\documentclass[pra,twocolumn,floatfix,notitlepage,superscriptaddress,longtable,nofootinbib]{revtex4-2}
\usepackage{amsmath}
\usepackage{lipsum} % for some text
\usepackage{verbatim}
\usepackage{epsfig}
\usepackage{subfigure}
\usepackage{graphicx}
\usepackage{enumitem}
\usepackage{amsfonts}
\usepackage[final]{showlabels}
\usepackage{enumitem}
\usepackage[figuresright]{rotating}
\usepackage{amssymb}
\usepackage{amsmath}
\usepackage{psfrag}
\usepackage{mathtools}
\usepackage[export]{adjustbox}
\usepackage{xr}
\usepackage{bm}% bold math
\usepackage[colorlinks,linkcolor=blue,anchorcolor=blue,citecolor=blue,urlcolor=blue]{hyperref}
\usepackage[version=4]{mhchem}
\usepackage[svgnames]{xcolor}
\def\be{\begin{equation}} \def\ee{\end{equation}}
\def\bea{\begin{eqnarray}} \def\eea{\end{eqnarray}}

\def\bpm{\begin{bmatrix}} \def\epm{\end{bmatrix}}

\newcommand{\bigO}{\mathcal{O}}

\newcommand{\highlight}{\color{black}}

\makeatletter
\newcommand*{\balancecolsandclearpage}{%
	\close@column@grid
	\clearpage
}
\makeatother

\begin{document}

\title{Consistent monitoring of quantum fluctuations}
\author{Xiangyu Cao}
\affiliation{Laboratoire de Physique de l'\'Ecole normale sup\'erieure, ENS, Universit\'e PSL, CNRS, Sorbonne Universit\'e, Universit\'e Paris Cit\'e, F-75005 Paris, France}
\date{\today}

\begin{abstract}
Recent works on the decoherent histories formalism suggested that slow-evolving macroscopic quantities (extensive sums of local observables) in quantum many-body systems can be consistently monitored: The existence of past measurements does not alter future outcome distribution. Here, for Gaussian weak measurements, we show that fluctuations of macroscopic quantities cannot be consistently monitored in general, in contrast to their intensive mean value. Exceptions include fluctuations at infinite temperature, at critical points, and in semiclassical systems. We analytically quantify non-consistency in terms of susceptibility, and obtain related results on entropy growth under noisy unitary.
\end{abstract}
\maketitle

\textit{Introduction}.--
 Macroscopic quantities, such as total energy, particle number and magnetization, fluctuate around their mean values in thermodynamic systems. Characterizing these fluctuations has been a central theme of statistical mechanics since Einstein's theory of Brownian motion~\cite{einstein-1906}. Seminal works by Onsager and Machlup~\cite{onsager31,onsager31-2,onsager-machlup,onsager-machlup2} described Gaussian fluctuations around thermal equilibrium as a stochastic process. The macroscopic fluctuation theory~\cite{mft-prl,mft-phystat,mft-review} provides the statistical law of fluctuations (including rare events) in weakly driven diffusive systems in terms of a variational principle; see Refs~\cite{kld-kpz,bettelheim-kmp,mallick-mft,grabsch-sfd} for recent advances. General laws of fluctuations~\cite{crooks-99,gallavotti-cohen,gallavoti-cohen-long,Jarzynski} are cornerstones of modern statistical physics.  %of ~\cite{}, and Jarzynski~\cite{} 

In classical systems, time-dependent fluctuations can be passively monitored~\cite{single-file-diffusion-ob,rna-crooks}. What about quantum systems~\cite{Bernard_2021}? Of course, quantum systems have fluctuations; they can also be continuously monitored~\cite{Wiseman_Milburn_2009}. The question is whether one may in principle probe the statistical law of the fluctuations \textit{consistently}, that is, without disturbing it~\cite{strasberg-scipost}. A textbook example of non-consistency is the double-slit experiment, where the interference pattern is disturbed by the existence of an earlier measurement, that is, if one detects through which slit the electron passes. % detector backreaction always exists in practice, it can be made arbitrarily small without reducing the measurement precision. 

The non-consistency is a signature of quantum coherence. When some fluctuating quantity can be monitored consistently, it is said to have decoherent (or consistent) histories~\cite{griffiths-coherent-histories,griffiths-93-prl,gellmann-hartle-prd,dowker-halliwell-prd,omnes-review,halliwell-review}, and considered to behave ``temporal-classically''~\cite{luppi2025temporal}. Recently, the emergence of decoherent histories in many-body quantum systems has attracted considerable attention ~\cite{gemmer-steigeweg-14,gemmer-16-markov,classicality-thermalization,strasberg-scipost,dhc-prx,wang-strasberg,ferte25,Strasberg:2026xwx}; see also \cite{strasberg-scipost,milz-markov19,non-markovian-milz,markov-pollock,strasberg-classicality-markov} for relation to Markovianity. These works suggest that macroscopic quantities in chaotic many-body systems have \textit{approximate} decoherent histories, namely, they can be monitored consistently up to a certain ``coarse-grained'' resolution~(see \cite{brukner-leggett-garg} for related work on macro-realism~\cite{leggett-garg}).  The aforementioned works focused on extensive quantities that almost commute with the Hamiltonian and thus evolve slowly. Meanwhile, concerning more generic extensive observables, some basic quantitative questions are left unsettled. In particular, it is unknown whether one may consistently monitor thermal fluctuations at equilibrium, whose magnitude scales as the square root of the volume [$\bigO(\sqrt{V})$] in general.

In this Letter, {\highlight we address this question in a framework of Gaussian weak measurements~\cite{ferte25}, which is conceptually akin to the consistent histories formalism, despite technical differences.} We show that it is in general impossible to monitor consistently $\bigO(\sqrt{V})$ fluctuations. The disturbance is quantified by the linear response function (susceptibility). As a consequence, equilibrium thermal fluctuations cannot be monitored consistently in general, with two notable exceptions: at infinite temperature (where susceptibility vanishes), or at a critical point with $\gg \sqrt{V}$ fluctuations. To reach these conclusions, we analytically characterize a setup in which a short-range correlated many-body system interacts with a number of measurement ancillas at multiple time steps. %(this setup also occurs in works on quantum dynamical entropies,~\cite{Jorge,cao2025planckian,AF}). % ( which we shall comment on.)

 % Also, the extensive $\bigO(V)$ mean value can be consistently monitored; it does not fluctuate at equilibrium but can have nontrivial evolution after a non-equilibrium quench. This situation was considered in previous works~\cite{classicality-thermalization,dhc-prx}, and our findings here do not contradict theirs. 

\textit{Macroscopic quantity and Gaussianity.--} We consider a quantum many-body system defined on a $d$-dimensional lattice with $V \gg 1$ sites. A macroscopic quantity is an operator that is a sum over lattice sites: 
\begin{equation} \label{eq:Q}
	Q := \sum_{r} Q_r
\end{equation}
where the sum is over all lattice sites (or an extensive fraction thereof), and $Q_r$ is a local operator that acts on a neighborhood of the lattice site $r$. We assume that the system is initialized in a state $\rho$ with short-range correlation and evolves under a local Hamiltonian $H$. Then, 
the rescaled fluctuating part of $Q$, defined as
\begin{equation} \label{eq:q-def}
	q_t := \left( Q(t) -  \mathrm{Tr} [ \rho Q(t) ]  \right) / \sqrt{V},
\end{equation}
where $Q(t) := e^{iHt} Q e^{-iHt}$, satisfies a Wick theorem in the thermodynamic $V \to \infty$ limit: $n$-point correlation functions of $q_t$ factorize into two point correlations~\cite{cao2025planckian} (see Appendix). In particular, $q_t$ has order-unity fluctuation, $\mathrm{Tr} [ \rho \,  q_t^2 ]= \bigO(1)$ as $V \to \infty$.  In other words, we may treat $q_t$ \textit{effectively} as a linear combination of ladder operators in a free boson system, and $\rho$ as a Gaussian state. This emergent Gaussianity generalizes one of Onsager and Machlup's main assumptions~\cite{onsager-machlup}. It can understood of as a central limit theorem, or a consequence of cluster decomposition, and will allow us to obtain analytical results exact in the thermodynamic limit. No microscopic integrability is assumed; our results do not distinguish chaotic and integrable systems~\cite{wang-strasberg}.
 
\textit{Measurement model.--} To monitor the fluctuations, we need non-demolition (weak) measurements. They are performed by letting the system interact with ancillas and then projectively measuring the latter. Here, we consider a simple ancilla model, made of a quantum harmonic oscillator; this preserves Gaussianity of the whole setup. We initialize it in the ground state $| 0 \rangle$ of $h = (p^2 + x^2) / 2$, $[x,p] = i$, and let it interact with the system under the unitary 
\begin{equation}  \label{eq:Ut}
	U_t = e^{-i \gamma_t \, p q_t}, \gamma_t > 0. 
\end{equation}
Then we projectively measure the position $x$ of the oscillator. Since the position has a Gaussian vacuum fluctuation of variance $\left< 0 |  x^2 | 0 \right> = 1 / 2$, and $U_t$ displaces the position by $\gamma_t q_t$, we obtain a weak measurement of $q_t$ with imprecision $\sim 1/\gamma_t$, described by the Kraus operators 
\begin{equation}\label{eq:Kraus}
	K_{m} = \gamma_t^{\frac12} \pi^{-\frac14}  e^{- \gamma_t^2 (m - q_t)^2 / 2 }, m := x / \gamma_t \in \mathbb{R}.
\end{equation}
 $\gamma_t$ is the measurement strength. We shall let $\gamma_t = \bigO(1)$ in the $V\to \infty$ limit, so as to measure $q_t$ with nonzero precision; if $\gamma_t \ll 1$, the outcome would be dominated by the apparatus noise and contain vanishing amount of information about $q_t$. %Note also that we measure the macroscopic quantity as a single operator; we do not perform local measurements and sum up the outcomes.  %Note also that we measure the macroscopic quantity as a single operator;   %Our use of weak measurements \eqref{eq:Kraus} differs from 

There is another way of viewing \eqref{eq:Ut}: The momentum $p$ also has a Gaussian vacuum fluctuation $\left< 0 |  p^2 | 0 \right> = 1  / 2$, so \eqref{eq:Ut} acts on the system as a random unitary. (See~\cite{zurek-deco-review} for a similar property of the qubit \texttt{CNOT} gate.) A more precise measurement (larger $\gamma_t$) induces a stronger backreaction on the system, altering further the statistics of later measurements. 

We remark that, increasing $\gamma_t$ to  $\lambda \gamma_t$ ($\lambda > 1$) is equivalent to replacing $| 0 \rangle$ by a squeezed state $| 0_\lambda \rangle$ such that $ \langle 0_\lambda | x^2  | 0_\lambda \rangle = 1 / ( 2\lambda^2)$ and $ \langle 0_\lambda | p^2  | 0_\lambda \rangle= \lambda^2 / 2$. Also, a thermal initial state with $\left< x^2  \right> \left< p^2  \right> > 1/4$ would need a larger backreaction to achieve the equal measurement precision as the ground state. Thus, the ground state corresponds to an ideal detector allowed by the uncertainty principle, and we shall focus on that. 

\textit{Two-time setup.--} The minimal setup where we can show the impossibility of consistent monitoring consists of two measurements, one of $q_0$ followed by that of $q_t$. This is a many-body analogue of the double slit experiment. We will denote the ancillas' position and momentum operators by $x_0, p_0, x_t, p_t$, respectively. The perturbation $U_0$ will evolve $q_t$ to the following,
\begin{equation}
   \tilde{q}_t  := U_0^{-1}	q_t U_0 = q_t + i \gamma_0 p_0 [q_0, q_t] + \dots.
\end{equation}
Since $q_t$ and $q_0$ are sum of (effective) boson ladder operators, the commutator $ [q_0, q_t]$ is a $c$-number, and the remainder term $\dots$, which involves nested commutators like $[q_0, [q_0, q_t]]$, vanishes in the $V\to\infty$ limit. The ancilla at $t$ will measure $\tilde{q}_t$ instead of $q_t$; its position operator after the interaction is 
\begin{align}
	\tilde{x}_t :=& U_0^{-1} U_t^{-1} x_t U_t U_0 =  x_t +  \gamma_t \tilde{q}_t   \nonumber \\
	=&  x_t +  \gamma_t (q_t + i \gamma_0 p_0 [q_0, q_t]  ). 
\end{align}
The three terms above are independent Gaussian or constant, so we have 
\begin{equation} \label{eq:xt2-2}
	\left<  \tilde{x}_t^2 \right> = \frac{1}{2} +  
	\gamma_t^2 \left( C(t, t) + \frac{ \gamma_0^2}{2} \chi(t, 0)^2 \right), 
\end{equation}
where 
\begin{align}
	&\chi(t, s) = \theta(t-s) \left< i [q_s, q_t] \right> , C(t, s)  = \frac12 \left< \{q_t, q_s\} \right>  
\end{align}
 are the linear response and Keldysh correlation function, respectively, and $\left< [\dots] \right>$  denotes an average in the initial state of system and ancillas, $\rho \otimes (|0 \rangle \langle 0 |)^{\otimes 2}$.  Noting that $ \tilde{x}_0  = x_0 +  \gamma_0 q_0$, we also find that,
 \begin{equation} \label{eq:cov-t0}
 	\left<  \tilde{x}_t   \tilde{x}_0  \right>  = \gamma_t \gamma_0 C(t,0), 	\left<  \tilde{x}_0  \tilde{x}_0  \right>  = \frac12 +  \gamma_0^2 C(0,0),
 \end{equation}
 which, together with \eqref{eq:xt2-2}, fully characterize the joint outcome distribution as centered Gaussian.
 
Eq.~\eqref{eq:xt2-2} is our first main result. It implies that, as long as $\chi(t, 0) \ne 0$, consistent monitoring is impossible: The distribution of the measurement outcome at $t$ is modified by the presence of the earlier measurement. Had it not taken place (or been very weak), we would have $\gamma_0 \to 0$, and $\left<  \tilde{x}_t^2  \right> \to  1/ 2 +   \gamma_t^2 C(t,t)^2$. The difference with \eqref{eq:xt2-2}, $\propto \gamma_0^2$, is small only as $\gamma_0 \to 0$, that is, as the measurement at $t = 0$ becomes weak. Note that the change is caused by the mere interaction with the ancilla at $t = 0$; the ancilla need not be measured.   %The outcome correlations in eq.~\eqref{eq:cov-t0} are not affected by prior measurements,  which do not exist. 

 %

%We may quantify the overall monitoring error by comparing the variance of $\tilde{x}_0/\gamma_0, \tilde{x}_t / \gamma_t$ with that of $q_0, q_t$, respectively, 
%\begin{align}
%	\mathcal{E} &:= (C(0, 0) - \left< (\tilde{x}_0/\gamma_0)^2 \right>)^2 +  (C(t, t) - \left< (\tilde{x}_t /\gamma_t )^2 \right>)^2  \nonumber \\ 
%&	= \frac{1}{4 \gamma_0^4} + \frac14 \left( \frac{1}{ \gamma_t^2} + \chi^2  {\gamma_0^2} \right)^2 > \frac{\chi^2}2, \label{eq:E}
%\end{align} 
%where $\chi := \chi(t, 0)$, for any $\gamma_0, \gamma_t$. Indeed, more precisely we measure $q_0$, we need to increase $\gamma_0$, which 

Fluctuations \textit{can} be consistently monitored if $\chi(t, 0) = 0$. This happens when $\rho \propto \mathbf{1}$ is the maximally mixed state. Therefore, in general, thermal fluctuations can be consistently monitored at infinite temperature but not at finite temperature.  

$\chi$ is also small in semiclassical systems, for example a lattice of spin-$S$ with $S \gg 1$. The rescaled spin operators $s_r^{x, y, z} = S^{x,y,z}_r / S$ will have small commutators, $[s^x_r, s^y_r] = i s^z_r / S$, and so on. Thus, if $Q_r$ in \eqref{eq:Q} is a function of $s^{x,y,z}_r$, we will have $\chi \sim [q_t, q_s] = \bigO(1/S)$, while $C \sim \{q_t, q_s\}$ is still of order $1$. Hence, fluctuations in semiclassical systems can be consistently monitored.
%which is well-defined if all lattice sites have a finite-dimensional Hilbert space).

Finally, consistent monitoring is possible when $\rho$ is a (quantum or thermal) critical state with long-range correlations. Then, $Q$ can have large fluctuations of amplitude $\gg \sqrt{V}$. Indeed, suppose that $Q = \sum_r Q_r$, $\left< Q_r \right> = 0$ and the two-point correlation decays algebraically
\begin{equation} \label{eq:QQ-crit}
\left< Q_r  Q_{r'}\right> \sim |r - r'|^{-2 \Delta} 
\end{equation}
($\Delta$ is known as the scaling dimension), and that $\Delta < d/2$, where $d$ is the spatial dimension. Then the standard deviation of $Q$ scales as follows,
\begin{equation}  \label{eq:Q-std-crit}
{\highlight \sqrt{\left< Q^2 \right>} \sim L^{d- \Delta} \sim V^{1 - \Delta / d} \gg \sqrt{V},}
\end{equation}
where $L \sim V^{\frac1d}$ is the linear system size (see Appendix for a derivation). In particular, the case $\Delta = 0$ includes superposition/mixture of distinct macro-states at a first-order transition. Such long-range correlations go beyond the scope of our analytical method. However, we argue heuristically that the large fluctuations can be consistently monitored.  Because they can be probed by much weaker measurements ($\gamma_{0, t} \ll 1$), which in turn have negligible backreaction. When $\Delta > d/2$, the fluctuations have $ \bigO (\sqrt{V}) $ amplitude again, and consistent monitoring is not expected in general (see however \cite{ferte25} for an exception).  %[see however~\cite{ferte25} fo]. 

\begin{figure}
	\centering
	\includegraphics[width=\columnwidth]{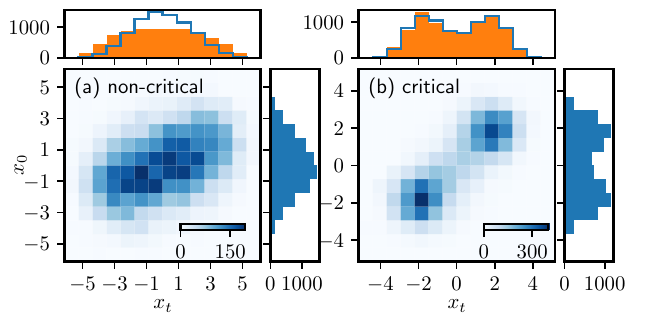}
	\caption{Joint and marginal outcome distributions from two-time ($t = 0,1$) monitoring of (rescaled) total magnetization in the quantum Ising model~\eqref{eq:ising} ($L = 16$) initialized in the ground state. The $t=0$ marginal histogram is plotted without filling color together with the $t = 1$ marginal in the top panels. We simulated $8000$ measurements for each data set. \textbf{(a)} Away from criticality ($J = 2/3 < 1$), the Gaussian fluctuations scaling as $\sim L^{1/2}$ cannot be consistently monitored: The measurement at $t = 0$ alters the outcome distribution at $t = 1$. See also Fig.~\ref{fig:diffs}-(a). $\gamma_0 = \gamma_1 = 1$.  \textbf{(b)} At quantum criticality ($J = 1$), the large critical fluctuations  rescaled accordingly {\highlight $q := \sum_j Z_j / L^{7/8}$} can be consistently monitored, $\gamma_0 = \gamma_1 = 2$.
	} \label{fig:hists}
\end{figure}
\textit{Numerics.}-- We test the above findings with finite-size exact numerics. For simplicity, we consider the 1D quantum Ising model. Its Hamiltonian is 
\begin{equation}\label{eq:ising}
H = \sum_{j = 1}^L (X_j + J Z_j Z_{j+1}) ,
\end{equation}
 where $X_j$ and $Z_j$ are Pauli matrices; we assume periodic boundary conditions. We first let $J < 1$, and initialize the system in its unique and short-range correlated ground state, $\rho = | \Psi \rangle \langle \Psi |$.  We simulated many experimental runs where the rescaled total magnetization $\sum_j Z_j / \sqrt{L}$ is measured at $t = 0$ and $t = 1$, with $\gamma_0 = \gamma_1 = 1$. The joint outcome distribution and the marginals are displayed in Fig.~\ref{fig:hists}-(a). The marginals are centered Gaussians to a good approximation, yet with different variances, $\left< \tilde{x}_1^2 \right> > \left< \tilde{x}_0^2 \right>$. Since $\rho$ does not evolve under $H$, the difference is solely due to the measurement backreaction, encoded by the last term of \eqref{eq:xt2-2}.  

Next, we repeat the simulation, but starting from the quantum critical ground state at $J = 1$. This is a scale-invariant state with long-range correlation in the (1+1)D Ising universality class. The total magnetization is known to have fluctuations of amplitude {\highlight $\sim L^{7/8}$} (the ``spin'' operator has $\Delta=1/8$), and we monitor the accordingly rescaled variable {\highlight$q := \sum_j Z_j / L^{7/8}$.} As a result, see Fig.~\ref{fig:hists}-(b), we find bimodal non-Gaussian distributions typical at the onset of symmetry breaking~\cite{delamotte}. Meanwhile, the two marginals are indistinguishable: The quantum critical fluctuations are consistently monitored. 

%\footnote{In general, in a critical system, if $Q$ is a sum over scaling operators with dimension $\Delta < d/2$ where $d$ is the space dimension, its fluctuation has amplitude $\sim \sim $, where $L$ is the linear size of the system. At (1+1)-d Ising critical point, the local magnetization is a scaling operator with $2nd = 1/8$.},

\begin{figure}
	\centering
	\includegraphics[width=\columnwidth]{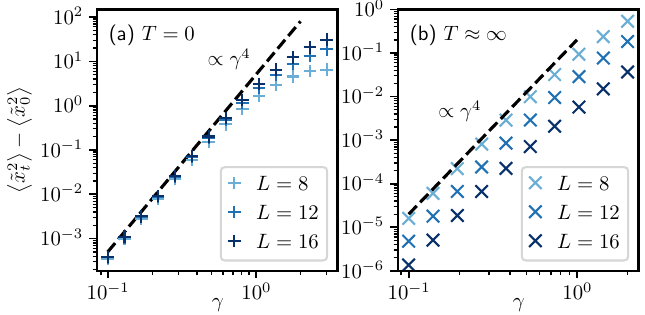}
	\caption{The difference between outcome variances $\delta := \left< \tilde{x}_t^2 \right> - \left< \tilde{x}_0^2 \right>$, in a two-time monitoring setup with the non-critical Ising model ($J = 2/3$), same as in Fig.~\ref{fig:hists}-(a), except that we vary the monitoring strength $\gamma_0 = \gamma_{1} = \gamma$ and system size. \textbf{(a)} Data with ground state initial condition are compared to the analytical prediction $\delta \propto \gamma^4$ (dashed line). \textbf{(b)}  
	The absolute difference $|\delta|$ averaged over $32$ Haar random initial conditions (per data point) tend to $0$ as $L$ increases.} \label{fig:diffs}
\end{figure}
Now we let $J < 1$ again, and analyze the variance difference $\delta := \left< \tilde{x}_t^2 \right> - \left< \tilde{x}_0^2 \right>$, $t = 1$, which quantifies the measurement disturbance in a stationary state. We compute $\delta$ numerically as a function of the monitoring strength $\gamma_0 = \gamma_1 =\gamma$, and for different system sizes $L$. In Fig.~\ref{fig:diffs}-(a), we show the result for the non-critical ground state. We find a power law dependence $ \delta \propto \gamma^4$, with an $L$-independent prefactor, in agreement with the analytical prediction \eqref{eq:xt2-2}, \eqref{eq:cov-t0}. This prediction is valid for  $\gamma \lesssim \gamma_*$ where $\gamma_*$ increases with $L$, indicating that our prediction is exact in the thermodynamic limit for any fixed $\gamma$. Upon replacing the initial state with Haar random pure states (which approximate the maximally mixed state), we find that $\delta$ rapidly tends to $0$ as $L$ increases for all values of $\gamma$, as expected, see Fig.~\ref{fig:diffs}-(b).

\textit{Multi-time setup.}-- To better appreciate the general structure underlying the two-time results, we consider a setup with $n$ successive measurements, happening at $\mathcal{T} = \{ t_1 < t_2 < \dots < t_n \}$. The measurement at $t$ has strength $\gamma_t$, and its ancilla has position and momentum operators $x_t, p_t$. With similar steps as above, we have
\begin{equation} \label{eq:xtildet}
\tilde{x}_t = x_t +  \gamma_t  \left( q_t + \sum_{u < t} 
	   i \gamma_u p_u [q_u, q_t]  \right),
\end{equation}
where the sum is over $\{u \in \mathcal{T}: u < t \}$. We then use \eqref{eq:xtildet} to calculate the correlations between $\tilde{x}_t$'s and $p_t$'s, the latter of which do not evolve as they commute with all $U$'s. We find that the covariance matrices $\mathbf{G}^{xx} := [	\left< \{ \tilde{x}_t, \tilde{x}_s \}  / 2\right> ]_{t,s \in \mathcal{T}}$, and similarly defined $ \mathbf{G}^{xp}$ and $  \mathbf{G}^{pp}$, are given by 
\begin{equation} \label{eq:main}
	\begin{bmatrix}
		\mathbf{G}^{xx}  & 	\mathbf{G}^{xp} \\ 
		\mathbf{G}^{px} & \mathbf{G}^{pp}  
	\end{bmatrix}
	= 
	\begin{bmatrix}
	\mathbf{I} / 2 +  \mathbf{C} + \mathbf{R} \mathbf{R}^T / 2 & \mathbf{R} \\
		\mathbf{R}^T & 		\mathbf{I} / 2
	\end{bmatrix},
\end{equation} 
where $\mathbf{I}$ is the $n \times n$ identity matrix and
\begin{equation}
	\mathbf{C}  := [\gamma_t   C(t,s) \gamma_s ]_{t,s \in \mathcal{T}}, \mathbf{R}  := [\gamma_t   \chi(t,s) \gamma_s ]_{t,s \in \mathcal{T}}. 
\end{equation}

Eq.~\eqref{eq:main} characterizes completely the state of the ancillas after interacting with the system. It is the main general result of this Letter, and worth some unpacking. First, the block $\mathbf{G}^{xx}$ unifies and generalizes \eqref{eq:xt2-2} and \eqref{eq:cov-t0} into the following, 
\begin{align}\label{eq:xx-gen}
&\left< \tilde{x}_t \tilde{x}_s \right> = \nonumber \\ 
 &\frac{\delta_{t,s}}2 + \gamma_t \gamma_s \left[C(t,s) + \frac{\gamma_u^2}2 \sum_{u} \chi(t, u) \chi(s, u)\right]. 
\end{align}
Viewing the ancillas as measurement apparatuses, \eqref{eq:xx-gen} determines the outcome distribution as an $n$-dimensional centered Gaussian. On the right hand side, the first term represents the apparatus noise. The second term is the intrinsic correlation between $q_t$ and $q_s$, encoded in the Keldysh correlation function. The last term is the perturbation brought by measurements prior to $t$ and $s$.  In general, the perturbation impacts multi-time correlations of the fluctuations as well as single-time marginals.
 
Eq.~\eqref{eq:xx-gen} can be viewed as an extension of the standard linear response theory, whose regime can be recovered by letting all $\gamma_t = \gamma \ll 1$. Then the third term can be viewed as a higher order ``post-linear response'' correction:
\begin{equation*}
	\left< \tilde{x}_t \tilde{x}_s \right> =  \frac{\delta_{t,s}}2 + \gamma^2 C(t,s) + \bigO(\gamma^4).
\end{equation*}
Meanwhile, the off diagonal block of \eqref{eq:main} does not have a post-linear response correction: 
\begin{equation} \label{eq:xp}
 \left<  \{ \tilde{x}_t,  \tilde{p}_s \} /2 \right>  = \gamma_t \gamma_s \chi(t, s). 
\end{equation}
In other words, the correlation between an early-time ancilla momentum and a later-time ancilla position encodes precisely linear response. This is intuitively reasonable, since the momentum sources the perturbation, whose effect is measured by the position. We stress that \eqref{eq:xx-gen} and \eqref{eq:xp} do not assume $\gamma_t \ll 1$; their exactness comes from the emergent Gaussianity in the thermodynamic limit.

\textit{Dynamical entropy}.-- A neat application of \eqref{eq:main} is to compute the Renyi and von Neumann entropies of the ancillas, 
\begin{equation} \label{eq:entropydef}
	\mathcal{S}_m : = (1-m)^{-1} \ln \mathrm{Tr}[\rho_A^m], \; \mathcal{S} := \lim_{m \to 1} \mathcal{S}_m,
\end{equation}
where $\rho_A$ is the ancillas' density matrix after interacting with the system. These entropies are closely related to the work of  on the entropy growth of a quantum system coupled to a noisy bath. Indeed, the ancillas constitute the bath, and their entropy equals that of the system \textit{if} the initial state $\rho$ is pure (our general results do not assume $\rho$ to be pure). $\mathcal{S}$ is also essentially the Alicki-Fannes quantum dynamical entropy~\cite{AF,Prosen_2007}. 

We can compute $\mathcal{S}_m$ and $\mathcal{S} $ because $\rho_A$ is a Gaussian state completely characterized by \eqref{eq:main}.  We find that the second Renyi entropy has a simple expression~\cite{Jorge},
\begin{equation} \label{eq:S-2}
	\mathcal{S}_2  = \frac12 \mathrm{tr} \ln  ( \mathbf{I} + 2 \mathbf{C}), 
\end{equation}
where the trace is over an $n$-dimensional space. All other entropies also depend only on the matrix $ \mathbf{C}$, See End Matter for explicit formulas~\eqref{eq:S-Renyi},\eqref{eq:S-vn} and their derivation.  Intriguingly, the entropies do not depend on the susceptibility $\chi$, which however appears as soon as we view the ancillas as measurement devices: $\chi$ affects the outcome distribution, and also measurement-induced purification~\cite{cao2025planckian}. In a sense, the ancillas keep a hidden record of the ``intrinsic'' fluctuation statistics, which can be revealed by a canonical transform [see \eqref{eq:canonical} in Appendix]. 

 %This is in contrast with other quantities that treating the ancillas more ``classically'', as measurement apparatus, such as the outcome 

\textit{Discussion.}-- We showed that $\bigO(\sqrt{V})$ fluctuations in many-body quantum systems cannot be consistently monitored in general. Measurements that can probe such fluctuations must be fine-grained enough to modify them. The non-consistency is a feature of quantum many-body systems away from any semiclassical limit, and is fundamentally different from the detector reaction in classical physics. The latter always exists in practice, but can be arbitrarily reduced without sacrificing measurement precision. Remark also that the intensive mean value of a macroscopic quantity, $Q/V$, can be monitored by much coarser measurements, with $\gamma_t \sim 1/\sqrt{V}$, and thus consistently in the thermodynamic limit~\cite{halliwell-hydro,classicality-thermalization}.

{\highlight Our exact results relied on two kinds of Gaussianity: the intrinsic Gaussianity of the observable, which we showed under general assumptions, and that of the measurement scheme~\eqref{eq:Kraus}, which we chose for solvability. It will be interesting to see to what extent our results apply to other measurement schemes.  In particular, the case of \textit{projective} measurements is directly relevant for the standard decoherent histories formalism.} An example is to monitor the sign of a macroscopic quantity $Q$ (assuming $0$ is not an eigenvalue of $Q$)~\cite{wang-strasberg,dhc-prx}. Because the sign changes abruptly at $Q = 0$, the measurement coarseness is not obvious to determine \textit{a priori}, and can be affected by how $Q$ fluctuates in a given state. If $Q = \pm V (c + o(1))$, as in a macroscopic superposition, we will monitor essentially the intensive quantity $Q / V$, which can be done consistently~\cite{dhc-prx}. Yet, if $Q \sim  \pm \sqrt{V}$, per our results, we expect the sign measurement to be sufficiently fine-grained to induce re-coherence~\cite{Strasberg:2026xwx}. Finally, the possibility (in principle) of resolving $Q \sim \pm \bigO(1)$ may result in more delicate microscopic coherent effects~\cite{wang-strasberg}.  {\highlight In summary, our results suggest a rather general failure of consistent histories for fast-evolving macroscopic quantities; explicitly checking this claim will be left to future work.}
%{\highlight In summary, our results indicate }

From a quantum information viewpoint, many-body quantum fluctuations seem to be a private random source that reveals passive eavesdroppers~\cite{ekert}.  It will be interesting to see how to use them as a resource for quantum cryptography~\cite{garratt2025private}.

\textit{Data availability}. The data that support the findings of this article are openly available ~\cite{data}.

\begin{acknowledgments}
	I thank Denis Bernard, Jorge Kurchan and Max McGinley for helpful discussions. I thank Denis Bernard, Antoine Tilloy, and the Anonymous Referees for valuable feedback on the manuscript.
\end{acknowledgments}

\bibliography{refs.bib}

\newpage
\appendix

\section{Emergent Gaussianity}
For the reader's convenience, we recall the argument of \cite{cao2025planckian} showing the emergent Gaussianity of macroscopic quantities in the thermodynamic limit. 

Consider first $n$ macroscopic quantities $Q^i = \sum_r Q^{i}_r$ (the superscript is not an exponent here and below), $i = 1, \dots, n$, such that $Q^{i}_r$ are operators supported in a neighborhood of $r$. Without loss of generality, we assume $ \left< Q^{i}_r \right> = 0$, where $ \left<  [\dots] \right> = \mathrm{Tr}[\rho[\dots] ]$, so that the rescaled quantities are $q^{i} = Q^{i} / \sqrt{V}$.  

The key assumption leading to Gaussianity is  that $Q^{i}$ has short range correlation with respect to a state $\rho$, with correlation length $\xi$. It means that, if a set of $n$ points can be decomposed into $m$ \textit{clusters}, $\{ 1, \dots, n \} = \sqcup_{a=1}^m C_a$, such that $|r_i - r_j| \ge R$ if $i, j$ are not in the same cluster, the following $n$ point correlation function factorizes up to an exponentially small error term, 
\begin{equation} \label{eq:decomp}
	\left<  \prod_{i=1}^n Q^{i}_{r_i} \right> = \prod_{a=1}^m \left<    \prod_{i \in C_a} Q^{i}_{r_i} \right>  + \bigO(e^{- R / \xi}). 
\end{equation}
We shall choose $R$ so that $V^{\frac1d} \gg R \gg \xi$. Note that in each expectation value, the operators are ordered by $i$. For example, if $n = 4$ and the clusters are $\{1, 4\}, \{2, 3\}$, the above equation reads $ \left< Q^1_{r_1} Q^2_{r_2} Q^3_{r_3} Q^4_{r_4} \right> =  \left< Q^1_{r_1} Q^4_{r_4} \right> \left< Q^2_{r_2} Q^3_{r_3} \right> + \bigO(e^{- R / \xi})$.   (We also assume that all $k \le n$ point correlations' absolute value are uniformly bounded.)

We now compute the $n$ point correlation function of the rescaled quantities, 
\begin{align} \label{eq:multipoint}
	\left< \prod_{i} q^{i} \right> = 
V^{-\frac{n}2}	\sum_{r_1, \dots, r_n} \left< \prod_{i=1}^n Q^{i}_{r_i} \right>. 
\end{align}
The sum above can be written as a sum over partitions (cluster decompositions) , and configurations $r_1, \dots, r_n$ that satisfy the decomposition (but not any finer decomposition). Now, in the $V\to\infty$ limit, the number of $m$-cluster configurations scales as $V^m$, since the points in the same cluster are bound to one another by an $\bigO(R)$ distance. Hence any decomposition with $m < n  / 2$ clusters will have a vanishing $\sim V^{m - \frac{n}2 } \ll 1$ contribution to \eqref{eq:multipoint}. Moreover, by \eqref{eq:decomp} and the assumption $\left< Q^{i}_r \right> = 0$, a decomposition with a one-point cluster also has vanishing $\bigO(e^{- R / \xi})$ contribution to \eqref{eq:multipoint}. Thus, the only surviving decompositions have $m = n/2$ clusters of size $2$ (so $n$ must be even). This shows that  $\left< \prod_{i} q^{i} \right>$ satisfies an approximate Wick theorem, that is, it factorizes into products of two point correlations, up to $\bigO(e^{-R / \xi}, V^{-1/2})$ error, which can be made arbitrarily small in the $V\to\infty$ limit. 

We may apply the above argument to the case where $Q^i_r= Q_r(t_i)$. The assumption \eqref{eq:decomp} remains true if we fix $t_1, \dots, t_n$ while taking the $V\to \infty$ limit, and replace $\xi$ by $\xi + v \max(t_i)$ where $v$ is the Lieb-Robinson velocity~\cite{LRbound}. This results in the multi-time approximate Gaussianity that we use in the main text.

 \section{Scaling at criticality}
{\highlight We provide the derivation of \eqref{eq:Q-std-crit} from \eqref{eq:QQ-crit}, and the assumption $\Delta < d / 2$. Indeed, we have 
\begin{align}
	\left< Q^2 \right> &= \sum_{r,r'} 	\left< Q_r Q_{r'} \right>   \nonumber  \\
&	 \sim \int \mathrm{d}^d r \mathrm{d}^d r' | r - r'|^{-2\Delta } \nonumber \\ 
&    \sim  L^d \int_{|y|< L}  \mathrm{d}^d y |y|^{-2\Delta }  \sim L^{2d - 2\Delta}. 
\end{align}
The first equality is by definition, the second line uses \eqref{eq:QQ-crit} and approximates a sum by an integral. Then we change variable to $y = r - r'$, and observe that the integral over $y$ is dominated by the large distance cutoff $y\sim L$ (since $\Delta < d / 2$).  Eq.~\eqref{eq:Q-std-crit} follows by taking the square root. When $\Delta > d / 2$, the integral will be dominated by a short-distance cutoff $|y| \sim 1$, and hence $ 	\left< Q^2 \right> \sim L^d  \sim V$ as in the non-critical case.}

\section{Ancilla entropies}
We calculate the Renyi entropies \eqref{eq:entropydef} of the ancillas. The idea is to bring the covariant matrix \eqref{eq:main} to a standard form, in order to reduce the calculation to that of single harmonic oscillators.  

To start, note that $\mathbf{C}$ is a positive semi-definite matrix, so can be diagonalized: $\mathbf{C} =  \mathbf{U} \mathbf{\Lambda} \mathbf{U}^{T}$ where $  \mathbf{U}  \in \mathrm{O}(n)$ and  $\mathbf{\Lambda} = \mathrm{diag}(\lambda_1, \dots, \lambda_n)$ is a diagonal matrix with eigenvalues of $\mathbf{C}$. Then, crucially, observe from \eqref{eq:main} that
\begin{equation}
	\begin{bmatrix}
		\mathbf{G}^{xx}  & 	\mathbf{G}^{xp} \\ 
		\mathbf{G}^{px} & \mathbf{G}^{pp}  
	\end{bmatrix} = 
	\frac12\mathbf{A}
	\begin{bmatrix}
		(	\mathbf{I}  +  2\mathbf{\Lambda} )^{\frac12}& 0 \\
		0 & 			(	\mathbf{I}  +  2\mathbf{\Lambda} )^{\frac12}
	\end{bmatrix}   \mathbf{A}^T
\end{equation}
where
\begin{equation} \label{eq:canonical}
	\mathbf{A} =	\begin{bmatrix}
		(	\mathbf{I}  +  2\mathbf{C} )^{\frac12} \mathbf{U} &  \\ 
		\mathbf{R} & 	(	\mathbf{I}  +  2\mathbf{C} )^{-\frac12}  \mathbf{U}. 
	\end{bmatrix}
\end{equation}
$ \mathbf{A} $ describes a canonical transform that turns the covariance matrix into a standard form, that of $n$ uncoupled harmonic oscillators of unit frequency, each at a thermal state with mean energy $\epsilon = (1 + 2 \lambda)^{\frac12} / 2$,  $\lambda = \lambda_1, \dots, \lambda_n$.  This canonical transform reveals the intrinsic fluctuation statistics hidden by the ancillas. 

Now we recall some basic facts about a harmonic oscillator. With frequency $\omega$ at inverse temperature $\beta$, the partition function is
$$ Z_\beta^{-1} = y- y^{-1}, y = e^{\beta \omega / 2}. $$ 
The dimensionless energy $\epsilon = E / \omega $ satisfies 
$$ 2 \epsilon =  \frac{y^2 + 1} {y^2 - 1},  y^2 = \frac{2 \epsilon + 1}{2 \epsilon - 1} = 
\frac{\sqrt{1 + 2 
\lambda } + 1}{ \sqrt{1 + 2 \lambda} - 1}. $$
This implies
$$ y^{\pm 1} =  (\sqrt{1 + 2 
\lambda } \pm 1)  / \sqrt{2\lambda}. $$
 The  $m$-th Renyi entropy $s_m$ satisfies
$$ e^{ (m-1) s_m } = \frac{ Z_\beta^m}{ Z_{m \beta} } = \frac{y^m  - y^{-m}}{(y - y^{-1})^m} = \lambda_+^m -  \lambda_-^m$$
where 
$$\lambda_{\pm} = \frac12 (\sqrt{1 + 2 \lambda } \pm 1)  = \epsilon \pm \frac12. $$
In particular, the von Neumann entropy is
\begin{align*}
s_{m \to 1} =& \lambda_+  \ln \lambda_+ - \lambda_-  \ln \lambda_-   \\
=& (\epsilon + \frac12 ) \ln  (\epsilon + \frac12 )  -  (\epsilon - \frac12 ) \ln  (\epsilon - \frac12 ) 
\end{align*}
and the second Renyi entropy satisfies
$$
e^{s_{2}} = \sqrt{1 + 2\lambda} = 2 \epsilon.
$$
The total Renyi entropies are obtained by summing  $s_m$ over $\lambda = \lambda_1, \dots, \lambda_n$: 
\begin{equation} \label{eq:S-Renyi}
	\mathcal{S}_m =\frac1{m-1}  \mathrm{tr} \ln  \left[ \mathbf{C}_+^m -  \mathbf{C}_-^m \right], 
\end{equation}
where 
$$ \mathbf{C}_{\pm} := \frac12 \left( \mathbf{I} + 2\mathbf{C} \right)^{\frac12} \pm  \frac{\mathbf{I}}2. $$
The von Neumann entropy is 
\begin{equation} \label{eq:S-vn}
	\mathcal{S} = \mathrm{tr} \left[ \mathbf{C}_+ \ln  \mathbf{C}_+ - \mathbf{C}_- \ln  \mathbf{C}_- \right], 	
\end{equation}

\textit{Entropy rates.--} Here we compare the entropies \eqref{eq:S-Renyi}-\eqref{eq:S-2} to purification, which is the average decrease of the system's entropy after the ancillas' positions are measured.  The rate of purification was calculated~\cite{cao2025planckian} with the same assumptions of this work, for thermal initial states, and in the continuum time limit. The latter is defined by taking $\mathcal{T} = \{  k \delta t: k = 1, \dots, n \}$ and $\gamma_t \equiv \sqrt{\delta t / 2} $ for all $t \in \mathcal{T}$ (we make this choice to match the normalization of \cite{cao2025planckian}). We shall also assume that the correlation functions are stationary, and denote their Fourier transform by the same symbol:
\begin{equation}
	\begin{pmatrix} C(t, s) \\ \chi(t, s) \end{pmatrix}  
	= \int \frac{d \omega}{2 \pi } e^{i (t-s) \omega}  \begin{pmatrix} C(\omega) \\ \chi(\omega) \end{pmatrix}  
\end{equation}
Then, \eqref{eq:S-2} implies the following entropy growth rate as $t = n \delta t \to \infty, \delta t \to 0$, 
\begin{equation} \label{eq:S2rate}
	\mathcal{S}_2 / t  \to   \int \frac{d \omega}{2 \pi}  \frac12 \ln (1 + C(\omega) ). 
\end{equation}
(Similar formulas can be found for other entropies.) Meanwhile, the von Neumann purification rate from a thermal initial state $\rho \propto e^{-\beta H}$  is~\cite{cao2025planckian}
\begin{equation}  \label{eq:Jrate}
	J / t \to  \int \frac{d \omega}{4 \pi} \frac{\beta \omega}{\sinh(\beta \omega)} \frac{C(\omega)}{1 + C(\omega) + |\chi(\omega)|^2/4} . 
\end{equation}
We may obtain the formula for the second Renyi purification rate by the following trick: Recall from \cite{cao2025planckian} that the factor $\beta \omega$ comes from $\partial_{\epsilon} s_1 = \beta \omega$. Replacing it with $s_2 = \ln (2 \epsilon) $, $\partial_\epsilon s_2  = 1/\epsilon =  2 \tanh(\beta\omega / 2)$, we obtain 
\begin{equation} \label{eq:J2rate}
J_2 / t \to  \int \frac{d \omega}{4 \pi} \frac{C(\omega) \mathrm{sech}(\beta\omega/2)^2}{1 + C(\omega) + |\chi(\omega)|^2/4} . 
\end{equation}

The entropy growth rate \eqref{eq:S2rate} and the purification rates \eqref{eq:Jrate}, \eqref{eq:J2rate} have two main differences. First, the former does not depend on susceptibility $\chi$, while the latter do. Second, the former can be positive even at zero temperature, while the latter must vanish as $\bigO(1/\beta)$ at low temperature.

%

%
%It follows that all the entropies can be computed explicitly in terms of the spectrum of $\mathbf{C}$: Remarkably, there is no dependence on susceptibility.
\end{document}